\documentclass[twocolumn,showpacs,amsmath,amssymb,aps,prd,floats,nofootinbib]{revtex4-2}
\usepackage{graphicx}
\usepackage[switch,pagewise,running]{lineno}
\usepackage{dcolumn}
\usepackage{bm}
\usepackage{xspace}
\usepackage{hyperref}
\hypersetup{colorlinks,linkcolor=blue,citecolor=blue,urlcolor=blue}  
\usepackage{color,xcolor}

\usepackage{booktabs} 

\renewcommand{\arraystretch}{1.3}

\begin{document}
\title{The Stochastic Dissipation Model for the Steady State Neutrino and Multi-Wavelength Emission of TXS~0506+056}
\author{Zhen-Jie Wang$^{1,2}$, Ruo-Yu Liu$^{1,2,3}$, Xiang-Yu Wang$^{1,2}$}
\affiliation{$^1$School of Astronomy and Space Science, Nanjing University, 210023 Nanjing, Jiangsu, China; \textcolor{blue}{ryliu@nju.edu.cn}\\
$^2$Key Laboratory of Modern Astronomy and Astrophysics (Nanjing University), Ministry of Education, Nanjing 210023, China\\
$^3$Tianfu Cosmic Ray Research Center,
 Chengdu 610000, Sichuan, China\\}

\begin{abstract}
    The blazar TXS~0506+056 has been suggested to be a potential high-energy neutrino source thanks to the observations of IceCube, which found outburst-like neutrino emissions during 2014-2015 and 2017 in the transient emission search, and a $3.5\sigma$ local significance in a 10-year time-integrated search. The conventional one-zone jet model cannot explain the observed neutrino flux during outbursts due to the constraint from the X-ray flux, leading to proposals of multi-zone models (e.g. two-zone model) with multiple radiation zones. In literature, it has been shown that multi-zone models may consistently explain the high-state neutrino emission and the multi-wavelength emission of TXS~0506+056, while the quasi-steady-state long-term emission has not been well studied. In this work, we investigate a physically based model for the quasi-steady-state neutrino and electromagnetic radiation under the same framework, and successfully reproduce the multi-messenger emission of TXS~0506+056. 
\end{abstract}

\maketitle

\section{Introduction}

    High-energy neutrinos are a good indicator of particle accelerators in the universe. The primary sources of high-energy neutrinos in the universe are still uncertain \citep{Aartsen_2013, IceCube_2013}. Blazars are considered potential sources of both high-energy cosmic-ray protons and neutrinos \citep{1995APh.....3..295M, Halzen_2002, Murase_2014PhRvD..90b3007M}. The photohadronic interaction model \citep{Mannheim_1992A&A...253L..21M, Mannheim_1993A&A...269...67M, 2018ApJ...864...84K, 2018MNRAS.480..192P, Padovani_2019MNRA, 2019ApJ...881...46R, Rodrigues_2019ApJ, Cerruti2019, IceCube:2021oqh} and the hadronuclear interaction model \citep{Dar_1997, 2018ApJ...866..109S, 2019PhRvD..99j3006B, Liu_2019, Xue_2022PhRvD.106j3021X} have been extensively studied to explain the blazar-neutrino association events. TXS 0506+056, a blazar with a redshift of $z=0.3365$ \citep{Paiano_2018}, is the first blazar to be significantly associated with neutrino events at a significance level above $3\sigma$. On 22 September 2017, IceCube detected a muon neutrino with an energy of 290 TeV \citep{2018Sci...361.1378I}, and about $13\pm5$ neutrinos were also detected by IceCube in 2014--2015 \citep{2018Sci...361..147I}. Some researchers have tried to explain the observed neutrinos with the standard one-zone model of the jet, but they found that the maximum expected neutrino event rate under the one-zone model is too small to explain the observations \citep[e.g.,][]{2018ApJ...864...84K, 2019NatAs...3...88G,Rodrigues_2019ApJ}, which has to be ascribed to Eddington bias or an upward fluctuation. However, a cascade neutrino event with an estimated energy of $224\pm75$\ TeV was detected by the new neutrino telescope Baikal-GVD in April 2021 \citep{Baikal-GVD_2021}, and then a $\sim$ 170 TeV neutrino, IC-220918A, was again detected in spatial coincidence with TXS~0506+056 by IceCube on 18 September 2022 \citep{IceCube_2022}. If these events are truly associated TXS~0506+056, Eddington bias cannot be a reasonable interpretation to the observed neutrinos. 

    The one-zone model assumes that the observed multi-wavelength emission comes from a compact spherical region, also known as a “blob”. In the framework of the one-zone model, the multi-wavelength emission of a blazar comes from one single blob in the jet. In general, the one-zone model is the most commonly used model and is successful in explaining the multi-wavelength SED of blazars \citep{Mastichiadis_1997A&A...320...19M, Bottcher_2002, Wang_2004, Finke_2008ApJ...686..181F, Zhang_2012ApJ...752..157Z, Ghisellini_2014Natur.515..376G, Yan_2014MNRAS.439.2933Y, Tramacere_2022A&A...658A.173T}. However, the one-zone model has some difficulties in explaining some observed properties of blazars. For example, in the framework of the one-zone model, the light curves (LCs) at different energy bands should be closely correlated \citep{Marscher_1985ApJ...298..114M, Hayashida_2015ApJ...807...79H, Zhang_2020PASJ...72...44Z}, but it has been observed that the patterns of variability in some blazars are not consistent between different wavelengths \citep{Liodakis_2019ApJ...880...32L, Petropoulou_2014A&A...571A..83P, MAGIC_Collaboration_2021A&A...655A..89M}. \cite{Liodakis_2019ApJ...880...32L} reported that some $\gamma$-ray flares and some optical flares are orphan events in some blazars, suggesting that the jet may have more complex properties than those described by the one-zone model. Meanwhile, the one-zone model can't produce the flat radio spectrum that is a common feature of blazars \citep{Planck_Collaboration_2011A&A...536A...1P}. This is because the synchrotron self-absorption effect is unavoidable in the one-zone model, so the produced radio spectrum is steeper than the observed spectrum. 

    To explain the neutrino emission from TXS~0506+056, multi-zone models, such as the two-zone model proposed by \cite{Liu_2019, Xue_2019, Xue_2021}, have been developed to explain the neutrino observations. For instance, \cite{Xue_2019} proposed a two-zone model for the neutrino event of IC-170922A. In their model, the inner zone is responsible for $\gamma$-ray and neutrino production via $p\gamma$ interactions, while the outer zone dominates optical/X-ray emission via synchrotron and inverse Compton processes. Our previous work \citep{Zhen-Jie_2024ApJ...962..142W} has also shown that the stochastic dissipation model (a multi-zone framework) successfully reproduces the high-state neutrino emission and multi-wavelength electromagnetic emission of TXS 0506+056 at different epochs. However, the quasi-steady state background multiwavelength radiation is fitted purely phenomenologically by a polynomial formula in that work. While the high-state emission is well explained, the jet properties of the quasi-steady state remains unconstrained. On the other hand, if neutrinos are produced at high states of the blazar, we should also expect neutrino emissions from the low state but just with a lower flux. 

    In this work, we aim to model the quasi-steady state multiwavelength emission and neutrino emission of TXS~0506+056 under the multi-zone framework with a physically driven method. Following prior studies, the quasi-steady state background neutrino and broadband electromagnetic (EM) radiations are contributed by many discrete small radiation zones in the jet. Both the leptonic emission processes and the hadronic emission processes will be considered in each radiation zone. In the jet, the spatial scale of the radiation zone (blob) increases with the distance $r$ from the black hole, while the magnetic field strength and the bulk Lorentz factor decrease with $r$. These radiation zones are generated by energy dissipation events such as magnetic reconnection events in the jet \citep{Begelman_1998ApJ, Giannios_2009MNRAG, McKinney_2009MNRA, Petropoulou_2016MNRA}. In our model the probability of a dissipation event occurring at a distance $r$ from the black hole is characterized by a function $p(r)\propto r^{-\alpha}$ \citep{Ruo_Yu_Liu_2023MNRAS.526.5054L}. Here, $r$ represents the distance from the supermassive black hole (SMBH), and $\alpha$ is the index that governs the spatial evolution of the dissipation probability. If we introduce proton injection at each dissipation zone, the radiation of the blazar may serve as a natural target for neutrino production via the photopion process. Indeed, the 10-year time-integrated search of IceCube found a $3.5\sigma$ local significance of high-energy neutrino signal from TXS~0506+056\cite{IceCube_NGC1068_2022Sci...378..538I}, which may represent the long-term quasi-steady-state neutrino emission of the blazar.
    
    The rest of the paper is structured as follows. The model description and the radiation calculation methods are described in Section 2. The result of the SED modeling are presented in Section 3. Finally, a summary of our results are presented in Section 4. Throughout the paper, $H_0= $69.6 km s$^{-1}$ Mpc$^{-1}$, $\Omega_{m}=0.29$ and $\Omega_{\Lambda}=0.71$ are assumed\citep{Bennett_2014}. 

\section{MODEL DESCRIPTION AND METHOD}

    A model called the stochastic dissipation model has been proposed in \cite{Ruo_Yu_Liu_2023MNRAS.526.5054L} and \cite{Wang_Ze-Rui_2022}. Due to the occurrence of energy dissipation events (such as magnetohydrodynamic instabilities \citep{Giannios_2006A&A, Nalewajko_10.1111}, magnetic reconnection events, etc.) in the jet, numerous small discrete radiation zones are generated in the jet. Most of these radiation zones produce weak radiation, the superposition of the emission from these radiation zones forms the quasi-steady state background radiation. When a strong energy dissipation event occurs in the jet, a flare or outburst event of the blazar would be produced, which we define as the high-state radiation. Consequently, the stochastic dissipation model can be characterized by two primary components: the quasi-steady state background radiation and the high-state radiation. The probability of an energy dissipation event occurring per unit time and per unit length is phenomenologically characterized by a function given by $p(r)\propto r^{-\alpha}$ \citep{Ruo_Yu_Liu_2023MNRAS.526.5054L}. 

    According to this model, each blob within the jet can accelerate particles and generate broadband electromagnetic radiation, spanning from radio waves to $\gamma$-ray. Neutrinos can be also generated if protons are also involved in the acceleration process. Both the leptonic and hadronic processes are considered in our model. For the radiation of electrons, we consider electron synchrotron radiation and inverse Compton (IC) scattering in each blob. The IC scattering includes two sources of target radiation field: the synchrotron radiation of electrons, also known as the synchrotron self-Compton (SSC) \citep{Harris_2006} process, and external radiation photons from other blobs or broad line region/dusty torus, also known as external Compton (EC) scattering. For the radiation of protons, we considered several mechanisms, including the photopion process, the Bethe–Heitler process, and proton synchrotron radiation. Additionally, we took into account the emission of electron-positron pairs that are generated in the electromagnetic (EM) cascades initiated by these processes. The absorption of $\gamma$-ray by the soft photons present in the radiation zone and the extragalactic background light (EBL; \cite{Dominguez_2011MNRAS.410.2556D}) is also considered. 

    In our model, we assume that the jet has a truncated conical structure and numerous emitting blobs are distributed along the jet. We may artificially divide the jet into many consecutive segments for convenience of numerical treatment. Each segment contains several emitting blobs, the number of which depends on the dissipation probability $p(r)$ along the jet. We assume that the each blob is of a spherical geometry characterized by a radius $R$ and is permeated by an entangled magnetic field $B(r)$ which is assumed to be homogeneous within each blob. 
    
    It is hypothesized that the jet's bulk Lorentz factor $\Gamma$ remains constant up to a distance of 0.1\,pc from the black hole \citep{Potter_2013MN}. Beyond this distance, it decreases gradually, reaching a Lorentz factor of 2 at a distance of 100\,pc. Additionally, We approximate the jet's Doppler factor $\delta$ to be equal to the jet's bulk Lorentz factor $\Gamma$. The Doppler factor for $r>0.1\ \rm pc$ can be expressed according to the formulations presented in \citep{Potter_2013MN}
    \begin{equation}
    \begin{aligned}
    \delta(r_{i}) = \rm \delta_{0}-\frac{\delta_{0}-2}{\rm log\left(\frac{100 \ pc}{0.1 \ pc}\right)} \rm log\left(\frac{r_{i}}{0.1 \ pc}\right),
    \end{aligned}
    \end{equation}
    where $\delta_{0}$ denotes the Doppler factor at 0.1 pc from the black hole. 
    
    In our model, $r_{0}$ is the position of the jet base, $r_{max}$ is the end of the jet and $r_{i}$ is the distance of the $i$th segment from the black hole. {The dissipation rate $\dot{N}$ of the entire jet is given by 
    \begin{equation}
    \begin{aligned}
    \dot{N}=N/T=\int_{r_{0}}^{r_{max}}p(r) dr,
    \end{aligned}
    \end{equation}
    where $N$ is the total number of blobs generated in a period of time $T$.}
    
    If we assume the jet has a truncated conical structure and the radius of a dissipation zone $R(r_{i})$ at certain distance $r_{i}$ from the black hole can be obtained by
    \begin{equation}
    \begin{aligned}
    R(r_{i}) = \kappa r_{i} \rm tan\theta,
    \end{aligned}
    \end{equation}
    where $\theta$ is the jet’s half-opening angle and $\kappa$ is the ratio of the blob's radius to the radius of it's segment. We assume that the diameter of the blobs inside each segment is one half of the length of each segment, so we have 
    \begin{equation}
    \begin{aligned}
    r_{i+1} = r_{i}(1 + 4\kappa \rm tan\theta),
    \end{aligned}
    \end{equation}
    and
    \begin{equation}
    \begin{aligned}
    i_{max} = \rm ln(r_{max}/r_{0})/ln(1 + 4\kappa \rm tan\theta).
    \end{aligned}
    \end{equation}
    If we assume that the magnetic luminosity is conserved along the jet \citep{O'Sullivan_2009MN,Sokolovsky_2010arXiv}, the magnetic field strength in the blob of the $i$th segment can be approximated as
    \begin{equation}
    \begin{aligned}
    B(r_{i}) = B(r_{0})\frac{R(r_{0})}{R(r_{i})},
    \end{aligned}
    \end{equation}
    

    The kinetic luminosity of nonthermal particles and magnetic field can be obtained by \citep{Celotti_2008}
    \begin{equation}
    L_{k,i} = \pi R^{2}\Gamma^{2}\beta cU_{i}, 
    \end{equation} 
    where $U_{i}$ is the energy density of nonthermal particles or magnetic field, here i = $\left\{\rm e ,\rm p, \rm B\right\}$, represents electrons, protons and magnetic field respectively. 

    \begin{figure*}[htbp]
    \centering
    \includegraphics[width=2\columnwidth]{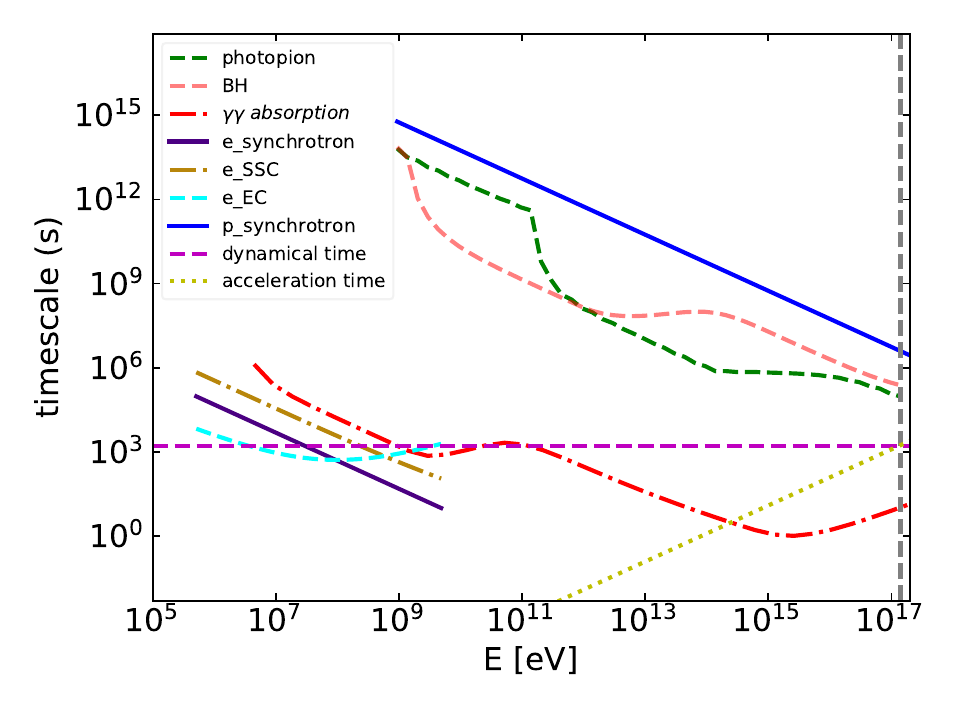}
    \caption{Relevant timescales in the comoving frame at the jet base, including cooling timescales of protons and electrons, and the interaction timescale of the internal $\gamma\gamma$ absorption within the radiation zone. The horizontal axis $\rm eV$ represents energy of particles in the comoving frame. The vertical grey dashed line shows the maximum proton energy allowed by the $t_{\rm acc} = $\rm min$ \left\{t_{\rm cool},t_{\rm p,dyn}\right\}$, {the $t_{\rm acc}$ is the acceleration time of proton, the $t_{\rm cool}$ is the cooling time of proton and $t_{\rm p,dyn}$ is the dynamical time of proton.}
    \label{1}}
    \end{figure*}

    In our model, the injection time $t_{\rm inj}$ of particles in each blob is assumed to be equal to the light crossing time $R(r_{i})/c$. The spectral distribution of electrons or protons is assumed to be a power-law function,
    \begin{equation}
    \begin{aligned}
    Q_{\rm e/p}(\gamma_{\rm e/p})&=\ Q_{\rm e/p,0}\gamma_{\rm e/p}^{-n_{\rm e/p}}, 
    \end{aligned}
    \end{equation}
    here $\gamma_{\rm e/p, min}\ \textless\ \gamma_{\rm e/p}\ \textless\ \gamma_{\rm e/p,max}$, $\gamma_{\rm e/p}$ is electron or protons Lorentz factor. Where $Q_{\rm e/p,0}$ is the normalization, and $n_{\rm e/p}$ represent the spectral index. In our model, $Q_{\rm e/p,0}$ can be obtained by the equation $\int Q_{\rm e/p}\gamma_{\rm e/p}m_{\rm e/p}c^{2}d\gamma_{\rm e/p} = L_{\rm e/p,inj}/\left(4/3\pi R^{3}\right)$. Here $L_{\rm e/p,inj}$ is the injection luminosity of electrons or protons, and $m_{\rm e/p}$ is the electron or proton rest mass. The temporal evolution of the electron (proton) distribution $N_{\rm e/p}(\gamma_{\rm e/p}, t, r_{i})$ in each blob can be obtained by the equation
    \begin{equation}
    \begin{aligned}
    \frac{ \partial N_{\rm e/p}(\gamma_{\rm e/p}, t, r_{i})}{\partial t} = &-\frac{ \partial }{\partial \gamma}[\dot{\gamma}(\gamma, t, r_{i})N_{\rm e/p}(\gamma_{\rm e/p}, t, r_{i}) ] 
        \\& - \frac{N_{\rm e/p}(\gamma_{\rm e/p}, t, r_{i})}{t_{esc}} + Q_{\rm e/p}(\gamma_{\rm e/p}),
    \end{aligned}
    \end{equation}
    here $\dot{\gamma}(\gamma, t, r_{i})$ is the total energy loss rate, and $t_{\rm esc} = \frac{10 R(i)}{c}$ is the escape time \citep{Gao_2017ApJ...843..109G}. Then $\dot{\gamma}(\gamma, t, r_{i})$ can be obtained by the equation
    \begin{equation}
    \begin{aligned}
    \dot{\gamma}(\gamma, t, r_{i}) = -\frac{\gamma}{t_{\rm dyn}} - \frac{\gamma}{t_{\rm rad}}
    \end{aligned}
    \end{equation}
    Here $t_{\rm dyn}\sim \frac{R(r_{i})}{c}$ is the dynamical timescale of each blob and is also the adiabatic loss timescale, and $t_{\rm rad}$ is the radiative cooling timescale.
    For the hadronic processes, the cooling time of protons can be obtained by $t_{\rm rad} = {\rm min}\left\{t_{p\gamma},t_{\rm BH},t_{\rm p,syn}\right\}$. {As shown in Figure~1, the $t_{p\gamma}$ is the cooling time of photopion process, the $t_{\rm BH}$ is the cooling time of Bethe–Heitler (BH) process and the $t_{\rm p,syn}$ is the cooling time of proton synchrotron radiation.} For the leptonic processes, $t_{\rm rad} = 3m_{\rm e}c / \left({4\left(U_{\rm B}+\kappa_{\rm KN}U_{\rm ph}\right)\sigma_{\rm T}\gamma_{\rm e}}\right)$. Here $U_{\rm B} = B^{2}/8\pi$ is the energy density of the magnetic field, $\kappa_{\rm KN}$ is a numerical factor accounting for the Klein-Nishina effect\citep{2005MNRAS.363..954M} and $U_{\rm ph}$ is the energy density of the soft photons. $U_{\rm ph}$ includes synchrotron photons from the blob itself, photons from the neighboring radiation zones and photons from the external regions such as broad-line region (BLR) or the dust torus (DT). $\sigma_{T}$ is the Thomson scattering cross section. If a certain segment reaches a quasi-steady state, the particle spectrum would present a cooling break, and the break energy (Lorentz factor) is determined by $t_{\rm dyn}=t_{\rm rad}(\gamma)$.

    In our model, we also considered the X-ray photons from corona surrounding the accretion disk \citep{Heckman_2014ARA&A}, the $\gamma$-ray emission of the blob can be absorbed by the X-ray photons if the blob is close to the black hole \citep{Righi_2019MNR}. We assume the emission of corona has a power-law spectrum 
    \citep{Ricci_2018MNR, Xue_2021}
    \begin{equation}
    L(E) = L_{\rm keV}(E/1 \ {\rm keV})^{1-\epsilon}, \ 0.1 \ {\rm keV} < E < 100 \ \rm keV,
    \end{equation}
    Then we assume $L_{\rm keV}$ is $5\times10^{43}\rm erg\ s^{-1}$ and its spectral index is $\epsilon = 1$ \citep{Xue_2021}. The energy density of photons from corona in the comoving frame can be estimated as
    \begin{equation}\label{eq:corona}
    u_{\rm corona} (r) = 
    \begin{cases} 
    \frac{\Gamma^{2}\int L(E)/EdE}{4\pi r_{\rm corona}^{2}c}, & r \leq 30r_{\rm Sch} \\
    0, & r > 30r_{\rm Sch} 
    \end{cases}
    \end{equation}
    { The scale of the corona is assumed to be $30r_{\rm Sch}$\citep{Ghisellini_2009}. For $r\leq30r_{\rm Sch}$, we assume $u_{\rm corona}$=$u_{\rm corona}(r=30r_{\rm Sch})$. For $r>30r_{\rm Sch}$, we simply assume $u_{\rm corona}$=$0$. This is because photons of corona almost radially propagate outward, and the scattering angle between protons and photons becomes small (or the target radiation field  is Doppler-deboosted in the comoving frame of blobs), leading to ineffcient $p\gamma$ interaction. Here, $r_{\rm Sch}$ is the Schwarzschild radius of the black hole, with $r_{\rm Sch}\sim 10^{14}\,(M_{\rm SMBH}/(3\times10^{8}M_{\odot})\,$cm\citep{Padovani_2019MNRA}.}
    
    We take BLR and DT radiation as an isotropic blackbody with a peak at $\approx 2 \times 10^{15}\Gamma \rm \ Hz$ \citep{2008MNRAS.386..945T} and $3\times10^{13}\Gamma \rm \ Hz$ \citep{2007ApJ...660..117C} in the jet comoving frame, respectively. The energy density of BLR and DT emission can be obtained by \citep{2012ApJ...754..114H}
    \begin{equation}
    u_{\rm BLR} = \frac{\eta_{\rm BLR}\Gamma^{2}L_{\rm d}}{3\pi r_{\rm BLR}^{2}c[1+\left(r/r_{\rm BLR}\right)^{3}]}, 
    \end{equation}
    and
    \begin{equation}
    u_{\rm DT} = \frac{\eta_{\rm DT}\Gamma^{2}L_{\rm d}}{3\pi r_{\rm DT}^{2}c[1+\left(r/r_{\rm DT}\right)^{4}]}, 
    \end{equation}
    where $\eta_{\rm BLR} = 0.1$ and $\eta_{\rm DT} = 0.1$ are the fractions of the disk luminosity $L_{\rm d}$ reprocessed into radiations of BLR and DT, respectively. $r_{\rm BLR} = 0.1\left(L_{\rm d}/10^{46}\rm erg \ s^{-1}\right)^{1/2} \rm pc$ and $r_{\rm DT} = 2.5\left(L_{\rm d}/10^{46}\rm erg \ s^{-1}\right)^{1/2}\rm pc$ are the characteristic distances of BLR and DT. We assume the BLR luminosity as $5\times10^{43}\rm erg\ s^{-1}$ \citep{Padovani_2019MNRA}, the disk luminosity $L_{\rm d}$ is assumed as $10~L_{\rm BLR}$, then the disk luminosity $L_{\rm d}$ is $5\times10^{44}\rm erg\ s^{-1}$ in our model, leading to $r_{\rm BLR}=0.022\,$pc and $r_{\rm DT}=0.56\,$pc. The photopion process and the BH process including the cascade are calculated following the method shown in \citep{2008PhRvD..78c4013K, Bottcher_2013}. {The method presented in \citep{2008PhRvD..78c4013K} is valid when Equations (19) of \cite{Bottcher_2013} are satisfied.} 

\section{SED Modeling and result} 

    \begin{figure*}[ht!]
    \centering
    \includegraphics[width=2\columnwidth]{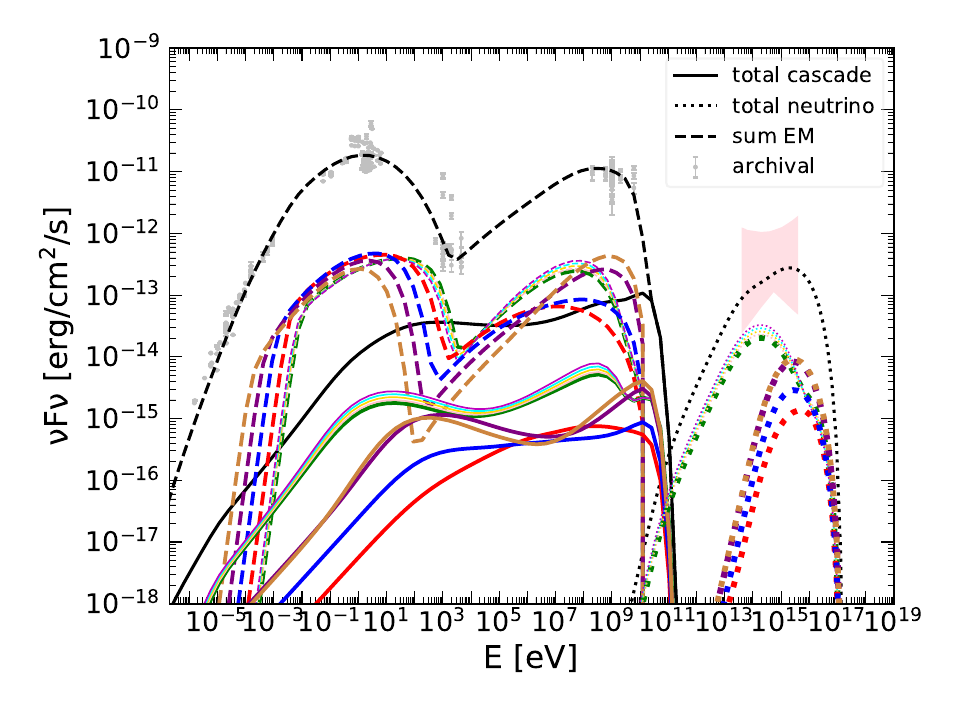}
    \caption{The quasi-steady state broadband SED of TXS~0506+056. The gray data points are historical archival data. The pink bow tie shows the muon-neutrino flux as measured in the quasi-steady state \citep{IceCube_NGC1068_2022Sci...378..538I}. The dashed, solid and dotted lines represent the leptonic radiation of the primary electrons, cascade radiation initiated by hadronic radiation processes and ($\nu_{\mu}+\bar{\nu}_\mu$) neutrino radiation, respectively. We also showcase the SEDs of 1st (0.0006 pc, green), 2nd (0.00067 pc, pink), 3rd (0.00074 pc, gold), 4th (0.00082 pc, cyan), 5th (0.0009 pc, magenta), 10th (0.0015 pc, red), 20th (0.004 pc, blue), {30th (0.01 pc, purple)} and 40th (0.03 pc, peru) segments in the jet. The solid and dotted black curve represent the total cascade radiation and neutrino radiation, and the dashed black curve represents the SED of the total EM radiation (including contribution from primary electrons and secondary electrons/positrons).}
    \label{2}
    \end{figure*}

    The SED fitting result are presented in Figure~\ref{2}. We see that the quasi-steady state multi-wavelength EM radiation and neutrino ($\nu_{\mu}+\bar{\nu}_\mu$) radiation of TXS~0506+056 can be reproduced by the stochastic dissipation model. The black dashed line represents the total EM radiation of all segments in the jet, the radiation processes include the leptonic radiation of the primary electrons and cascade radiation initiated by hadronic radiation processes. The black solid line represents the total cascade radiation initiated by hadronic radiation processes of all the segments in the jet. The black dotted line represents the total neutrino ($\nu_{\mu}+\bar{\nu}_\mu$) radiation. 

    The free parameters in our model are $r_{0}$, $r_{\rm max}$, $\kappa$, $\dot{N}$, $\theta$, $B_{0}$, $\alpha$, $\Gamma$, $L_{\rm p,inj}$, $L_{\rm e,inj}$, $\gamma_{\rm p,min}$, $\gamma_{\rm p,max}$, $\gamma_{\rm e,min}$, $\gamma_{\rm e,max}$, $n_{\rm e}$, $n_{\rm p}$. The SED modeling and result are shown in Figure~\ref{2}, the derived parameters are listed in Table 1. As shown in Table 1, the position of jet base from black hole is about $0.00055\rm\ pc$, which is approximately 17$r_{\rm Sch}$ with  $r_{\rm Sch}$ being the Schwarzschild radius. At this distance, high-energy protons in the blobs can effectively interact with the photons from the corona and the BLR, so the neutrino can be effectively produced by $p\gamma$ interactions at this location. 

\begin{figure*}[ht!]
    \includegraphics[width=2\columnwidth]{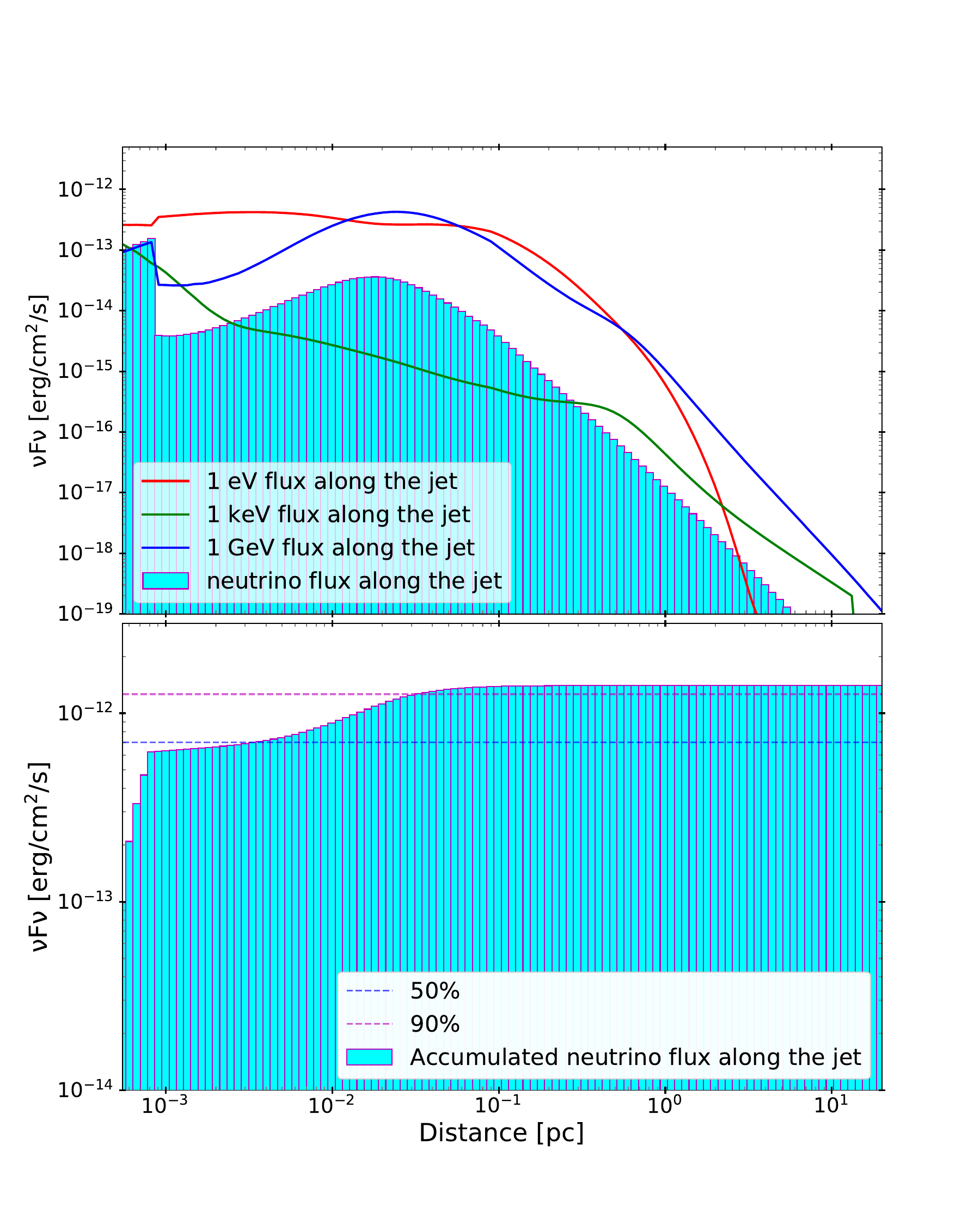}
    \caption{ The upper panel of the figure is the {integrated} neutrino ($\nu_{\mu}+\bar{\nu}_\mu$) flux {(10\,TeV-10\,PeV)} along the jet, and the lower panel of the figure is the accumulated {integrated} neutrino ($\nu_{\mu}+\bar{\nu}_\mu$) flux {(10\,TeV-10\,PeV)} along the jet. The horizontal axis represents the distance from black hole, the vertical axis represents the neutrino flux. The horizontal dashed lines represent $50\%$ and $90\%$ of the accumulated neutrino flux, respectively.
    }
    \label{3}
    \end{figure*}

    The top panel of Figure~\ref{3} shows the integrated neutrino ($\nu_{\mu}+\bar{\nu}_\mu$) flux ($\rm 10\,TeV-10\,PeV$) generated at each segment along the jet. {The neutrino flux peaks at the edge of the corona ($\sim$30$r_{\rm Sch}$). Within the corona, the neutrino flux distribution increases along the jet, followed by an abrupt drop at $\sim$30$r_{\rm Sch}$ due to our treatment of the corona radiation field (Eq.~\ref{eq:corona}). The neutrino production is dominated by BLR radiation field at larger radii peaking at $r_{\rm BLR}$. This spatial variation is primarily governed by the energy density of the soft photon field, and is also influenced by the dynamical timescale in each blob ($\propto r$). The latter is the reason why the neutrino flux distribution slowly rise with $r$ within $r_{\rm BLR}$. Infrared photons from DT does not contribute much to neutrinos. This is because the maximum proton energy, as determined by the dynamical timescale and acceleration timescale (e.g., Figure~\ref{1}), is about $10^{17}$\,eV (in the comoving frame). The typical photon energy of DT radiation is about 0.5\,eV in the comoving frame. Most of these photons fall below the energy threshold required for the photopion interactions and are therefore unable to participate in the photopion process.}

    
    The bottom panel of Figure~\ref{3} shows the accumulated {integrated} neutrino ($\nu_{\mu}+\bar{\nu}_\mu$) flux {(10 TeV-10 PeV)} along the jet, the accumulated neutrino flux represents the total neutrino flux between the local position and the black hole. The $90\%$ of the observed neutrino flux is contributed by regions within $\sim 0.03\rm \ pc$ of the jet. We may obtain the total neutrino ($\nu_{\mu}+\bar{\nu}_\mu$) event rate by $T_{\rm obs}\int_{\small 10\rm\ TeV}^{\small 10\rm\ PeV}(dN(E)/dEdt)A(E)dE$, where $A(E)$ \citep{IceCube_2020} is the effective area for this source and $T_{\rm obs}$ is the observation time. The obtained total neutrino event rate expected for IceCube (in unit of number per one year) in the quasi-steady state of TXS~0506+056 is about $0.2 \rm\ yr^{-1}$.  

    For reference, we also show the spatial distributions of optical (1\,eV), X-ray (1\,keV) and $\gamma$-ray fluxes (1\,GeV)  along the jet in the top panel of Figure~\ref{3}. We see that the optical and gamma-ray flux peaks at $\sim 0.01\,$pc from the black hole, while the X-ray flux peaks at the jet base. Given $B\propto r^{-1}$ and $U_{\rm BLR/DT}$ is constant at small distance, it is straightforward to understand why synchrotron X-ray flux distribution decreases as the distance and GeV gamma-ray flux (mainly results from EC scattering off BLR radiations) distribution peaks at $r\sim r_{\rm BLR}$. The optical emission is also from the synchrotron radiation of electrons, but the flux distribution peaks at larger distance. This is because the cooling of optical-emitting electrons (with energy of $\sim 100\,\rm MeV$) is dominated by the IC scattering off the X-ray corona's emission at the jet base (see the dashed cyan line in Figure~\ref{1}). 
        
    Based on the injection spectrum of both electrons and protons, as well as the magnetic field obtained in the fitting,{we may calculate the kinetic luminosity of the electrons, protons and magnetic field along the jet, and the maximum values are found to be $L_{\rm k,e}=8.6\times 10^{43}\,\rm erg~s^{-1}$, $L_{\rm k,p}=7.3\times10^{45}\,\rm erg~s^{-1}$ and $L_{\rm k,B}=3.6 \times 10^{45}\rm \,erg~s^{-1}$, respectively}. The mass of the SMBH of this source is suggested to be about $3 \times 10^8M_\odot$\cite{Padovani_2019MNRA}, so that the Eddington luminosity of the blazar is  $L_{\rm edd} \approx 4 \times 10^{46}(M_{\rm BH}/10^{8.5}M_\odot)\, \rm erg~s^{-1}$. The jet's total luminosity required in our model is below the Eddington limit, which is consistent with the steady state of the blazar that we deal with.
    
    It's worth noting that the obtained magnetic field strength is 90\,G at the jet base in our model. It is much higher than the magnetic field strength of the inner blob in the two-zone model \citep{Xue_2021} for the 2014-2015 outburst of TXS~0506+056 (i.e., 2\,G), which is located at $\sim 3r_{\rm sch}$ from the black hole . On the other hand, the proton kinetic luminosity obtained by \citet{Xue_2021} is several times higher than what we obtained here. Such differences are understandable because our model is for the steady state while \citet{Xue_2021} is for the high state. The low magnetic field found during the outburst therefore might be interpreted as the outcome that the magnetic energy at the jet base is dissipated into that of relativistic particles during the outburst. If so, it implies magnetic dissipation event (such as the magnetic reconnection) as the driven mechanism of the 2014-2015 outburst. 


    
    \begin{table*}[htbp]
    \normalsize 
    \renewcommand{\arraystretch}{1.35} 
    \setlength{\tabcolsep}{6pt} 
      \begin{center}
        {\small \caption{Summary of Model Parameters.}}
        \scalebox{1.2}{    
        \begin{tabular}{c c c} 
        \toprule  
        \hline
          \textbf{parameters} & \textbf{Values} & \textbf{Notes}\\
          \hline
          $r_{0}(pc)$ & 0.00055 & Position of jet base from black hole\\
          $r_{max}(pc)$ & 100 & Jet length \\
          $\kappa$ & 0.3 & Ratio of blob’s radius to its segment’s radius \\
          $\dot{N}$ & 0.4 & Blob generation rate of the entire jet \\
          $\theta(\circ)$ & 5 & Jet’s half-opening angle \\
          $B_{0}(G)$ & {90} & Magnetic field strength at jet base \\
          $\alpha$ & 1.8 & Index of the dissipation probability $p(r_{i})$ \\
          $\Gamma$ & 6.5 & Jet base’s bulk Lorentz factor \\
          $\rm L_{p,inj}(erg/s)$ & ${3.0\times10^{41}}$ & Injection proton luminosity in each blob \\
          $\rm L_{e,inj}(erg/s)$ & $2.0\times10^{40}$ & Injection electron luminosity in each blob\\
          $\gamma_{\rm p,min}$ & 1 & Minimum proton Lorentz factor \\
          $\gamma_{\rm p,max}$ & $8\times10^{7}$ & Maximum proton Lorentz factor \\
          $\gamma_{\rm e,min}$ & 1 & Minimum electron Lorentz factor \\
          $\gamma_{\rm e,max}$ & $10^{4}$ & Maximum electron Lorentz factor\\
          $n_{\rm e}$ & 1.7 & The spectral index of electron energy distribution \\
          $n_{\rm p}$ & 2.0 & The spectral index of proton energy distribution \\
          $L_{\rm k,e}$ & $8.6\times10^{43}$ & The kinetic luminosity of electrons \\
          $L_{\rm k,p}$ & ${7.3\times10^{45}}$ & The kinetic luminosity of protons \\
          $L_{\rm k,B}$ & ${3.6\times10^{45}}$ & The kinetic luminosity of magnetic field \\
          \bottomrule 
        \end{tabular}  }
    \end{center}
    \end{table*}
    
\section{Summary}

    In this work, we modeled the quasi-steady state neutrino and multi-wavelength emission of blazars under the stochastic dissipation model. By utilizing TXS~0506+056 as an example, we showed that the quasi-steady state neutrino ($\nu_{\mu}+\bar{\nu}_\mu$) and multi-wavelength emission of this blazar can be reproduced by the model. 
    
    As shown in Figure~\ref{2}, the quasi-steady state background radiation obtained by our model shows a “two-bump” morphology, which is dominated by leptonic radiation processes of the primary electrons. Similar to the one-zone model, the synchrotron radiation and the IC radiation are the dominant radiation mechanism for the low-energy bump and high-energy bump respectively, but the emissions are from all the blobs along the jet instead of a single blob. The radio spectrum can be also reproduced in this model. The spectrum of EM cascade initiated by interactions of protons roughly shows a $E^{-2}$ spectrum, which has a non-negligible contribution to the X-ray flux, where the `valley' between the two bumps in the SED occurs.
    Cooling of high-energy protons in the PeV energy regime is dominated by the photopion production. In this case, the 10-year long-term neutrino flux inferred from IceCube's observation can also be  explained. The neutrino ($\nu_{\mu}+\bar{\nu}_\mu$) event rate of one year in this case is about $0.2 \rm\ yr^{-1}$. During outburst or flare, the neutrino event could be higher given stronger injection rate of protons.  

    The main constraints on the neutrino ($\nu_{\mu}+\bar{\nu}_\mu$) flux come from the measured X-ray flux, because the overall EM spectrum would firstly saturate in the observed flux in the X-ray band, if we want to increase the neutrino flux by increasing either the proton injection luminosity or neutrino production efficiency. As shown in Figure~\ref{2}, the neutrino ($\nu_{\mu}+\bar{\nu}_\mu$) flux predicted by our model can explain the observed quasi-steady state neutrino ($\nu_{\mu}+\bar{\nu}_\mu$) flux, and the pair cascade emission doesn't saturate the X-ray emission at this time. 

    Finally, we would like to caveat that based on the current measurement by IceCube, we could not distinguish well whether the 10-year average neutrino ($\nu_{\mu}+\bar{\nu}_\mu$) flux of TXS~0506+056 is truly contributed by a quasi-steady emission component or mainly by several discrete outbursts. Our model is based on the former scenario, and in the latter scenario our prediction may be understood as an upper limit of the quasi-steady emission component. The next-generation neutrino instrument of $\sim 10\,$km$^3$ volume such as HUNT\cite{Huang2024icrc.confE1080H}, IceCube-Gen2\cite{Aartsen2021JPhG...48f0501A}, and TRIDENT\citep{Ye2023NatAs...7.1497Y} may detect much more neutrino events from the source and distinguish the two scenarios.

\section*{acknowledgments}

    This work is supported by National Science Foundation of China under grants No.~12393852 and 12333006.

\vspace{5mm}
\bibliographystyle{apsrev}
\bibliography{references}


\newpage

\end{document}